\documentclass[twocolumn,showpacs,preprintnumbers,amsmath,amssymb]{revtex4}
\usepackage{graphicx}% Include figure files
\usepackage{dcolumn}% Align table columns on decimal point
\usepackage{bm}% bold math

\begin{document}

\preprint{APS/123-QED}

\title{In-situ diagnostics of the crystalline nature of single organic
nanocrystals by nonlinear microscopy}% Force line breaks with \\

\author{Sophie Brasselet, V\'eronique Le Floc'h, Fran\c{c}ois Treussart, Jean-Fran\c{c}ois Roch}
\author{Joseph Zyss}
\affiliation{Laboratoire de Photonique Quantique et Mol\'eculaire, Ecole Normale Sup\'erieure de Cachan \\
61 avenue du Pr\'esident Wilson, 94235 Cachan Cedex, France }
\author{Estelle Botzung-Appert}
\author{Alain Ibanez}
\affiliation{Laboratoire de Cristallographie, BP 166, 38042
Grenoble Cedex 09, France }
\date{\today}

\begin{abstract}
We elucidate the crystalline nature and the three-dimensional
orientation of isolated organic nanocrystals embedded in a sol-gel
matrix, using a polarized nonlinear microscopy technique that
combines two-photon fluorescence and second harmonic generation.
This technique allows the distinction between mono-crystalline
structures and nano-scale poly-crystalline aggregates responsible
for incoherent second harmonic signals.
\end{abstract}


\maketitle

The optical properties of nanoparticles have recently attracted
much attention. In addition to metallic and semiconductor
nanoparticles which are now used as bio-markers and as the
building blocks of nanostructured materials \cite{NanopartBio},
their organic counterparts constitute an interesting alternative.
Advances in molecular engineering have enabled the design of
molecular structures of various resonances and symmetries with
optimized one- and two-photon absorption cross sections
\cite{2PhotonDyes}, or combining different optical properties such
as luminescence and second harmonic generation (SHG)
\cite{Blanchard}. In addition, macroscopic molecular arrangements
have been optimized using the tensorial oriented gas model
\cite{NLOcrystals}, which predicts that an enhancement of the SHG
efficiency is expected from the non-centrosymmetric crystalline
arrangement of efficient nonlinear molecules. Molecular
nanocrystals can be therefore envisioned as a new class of
multi-functional nano-scale materials. In the case of organic
nanocrystals however, the traditional crystalline characterization
techniques have raised many practical barriers due to their low
concentration and fragility. Consequently, the elucidation of
their crystalline nature has been so far indirect or averaged over
a large number of nanocrystals \cite{Vallee}.

In this letter, we show that two-photon nonlinear microscopy
permits in-situ characterization of isolated organic nanocrystals
grown in an amorphous sol-gel matrix. The diagnostic is based on
polarization resolved two-photon excited fluorescence (TPF) and
SHG. TPF is an incoherent process allowed in centrosymmetric
media, which exhibits a specific anisotropy depending on the
medium symmetry. On the other hand, SHG is the signature of a
crystalline non-centrosymmetric phase in the sample, with a
sensitivity down to the nanometric scale
\cite{NLO-LB,PrasadNanoCrystals}. We show that the polarization
analysis of both TPF and SHG from nanocrystals allows the
unambiguous discrimination between isolated mono-crystalline and
poly-crystalline systems. Moreover, once a nanocrystal has been
identified as mono-crystalline, a detailed model for both TPF and
SHG polarization responses accounting for the unit-cell symmetry
allows the determination of its three-dimensional orientation
within the host matrix.

The organic nanocrystals that we investigate are based upon the
$\alpha$-((4'-methoxyphenyl)methylene)-4-nitro-benzene-acetonitrile)
molecule (CMONS), which exhibits efficient luminescence and
quadratic nonlinearity under two-photon excitation
\cite{Wang,Treussart,Sanz}. The bulk crystalline phases of such
crystals have three possible polymorphic forms, two being
non-centrosymmetric with a very similar unit-cell crystal geometry
(forms $(a)$ and $(b)$)\cite{Sanz,CMONScristallo,Oliver}. The
preparation of the organic nanocrystals in sol-gel glasses relies
on the control of the nucleation and growth kinetics of the dye
confined in the pores of the gel \cite{Sanz}. CMONS nanocrystals
were grown in 1:1
tetramethoxysilane(TMOS):methyltrimethoxysilane(MTMOS) matrices
with a controlled CMONS:alkoxydes molar fraction. The sols
containing the alkoxyde solvent, water and organic phases are
directly spin-coated on a glass substrate. The presence of
nanocrystals in the sol-gel film after spin-coating was
ascertained by measurement of the material melting point using
differential scanning calorimetry. Previous works on such
materials have shown through bulk spectroscopy characterizations
that the nanocrystals are in the predominant thermodynamically
stable $(a)$ form, with a luminescence maximum peaking around
500-580 nm \cite{Sanz}. The particle mean size, which depends on
the preparation parameters such as temperature and matrix
porosity, ranges from 20 to 100 nm. In this work, we focus on the
smallest particles, which are obtained using a CMONS:alkoxydes
molar fraction of $4\times10^{-3}$.

\begin{figure}[!h]
\includegraphics[width=7cm]{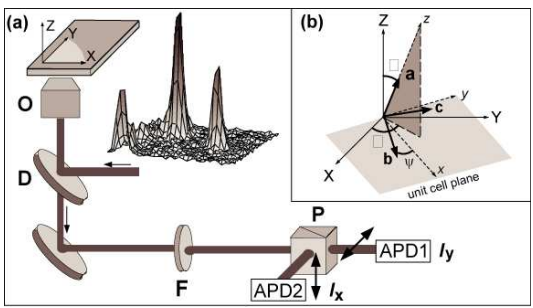}
\caption{\label{fig1} (a) Experimental detection scheme. F:
filter; D: dichroic mirror; P: polarizing beamsplitter; O:
microscope objective ($\times 60$, $N.A.$=$1.4$); APD1, APD2:
silicon avalanche photodiodes. The 3D plot shows a SHG raster scan
of a 7.6 $\mu$m$\times$7.6 $\mu$m area. Each spot is associated
with an isolated CMONS nanocrystal. (b) Orientation of the
unit-cell frame (a,b,c) in the macroscopic frame ($X,Y,Z$). The
$(x,y,z)$ axes relate to the nanocrystal frame with $(x,y)$
defining the unit-cell plane (see Fig.~\ref{fig2}).}
\end{figure}

CMONS nanocrystals immobilized in the sol-gel film (0.5 - 1 $\mu$m
thickness) are imaged using an inverted two-photon microscopy
set-up with an excitation wavelength of 987 nm from a Ti:Sa laser
(150 fs pulses duration, 82 MHz repetition rate) \cite{LeFloch}.
The reflection geometry, which uses a dichroic mirror to provide
the incident IR light onto the sample, is well suited for the
observation of SHG from structures of sub-wavelength size, this
optical process being in this case insensitive to nonlinear
phase-mismatch effects. The SHG signal, at half of the incident
wavelength (493.5 nm), is spectrally shifted from the TPF emission
maximum (at around 570 nm in the studied films), and can therefore
be detected separately using appropriate spectral filters
\footnote{Note that for all quantitative analysis, we account for
residual TPF emission in the SHG detection channel.}. TPF and SHG
scanning microscopy images exhibit isolated spots (about $1$ per
10 $\mu$m$^{2}$) with diffraction limited resolution
(Fig.~\ref{fig1}(a)). The polarized two-photon microscopy
technique consists of rotating the incident IR linear polarization
while recording the SHG or TPF signals along two perpendicular
($X$ and $Y$) polarization directions for each isolated
nanocrystal. We will show that this simple scheme allows the
distinction between mono-crystalline and poly-crystalline forms.

\begin{figure}[!h]
\includegraphics[width=6.5cm]{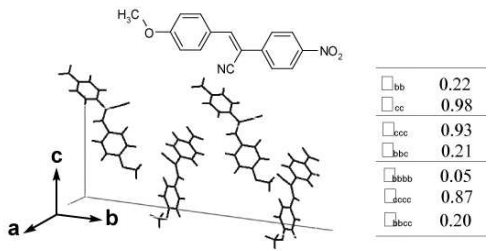}
\caption{\label{fig2} Molecular structure of CMONS and unit-cell
representation of the crystalline form (a), adapted from Ref.
\cite{CMONScristallo}. The unit-cell is constituted by four CMONS
dipoles tilted with an angle of $\pm25.5^{\circ}$ relative to the
(c) axis. The table shows the non-vanishing coefficients of the
microscopic susceptibility tensors of the CMONS unit-cell. All
out-of plane contributions involving the index (a) can be
neglected, as one can assume that the molecules lye in the (b,c)
plane of the unit-cell \cite{CMONScristallo}. The coefficients are
normalized so that their tensorial norm is equal to unity.
Non-diagonal coefficients are equal for permutations on all
indexes.}
\end{figure}

In order to model the polarization responses of both optical
processes, we first assume a mono-crystalline CMONS nanocrystal
whose unit-cell orientation is defined by the Euler set of angles
$\Omega=(\theta,\phi,\psi)$, as illustrated in Fig.~\ref{fig1}(b).
The CMONS crystalline unit-cell in its non-centrosymmetric form
$(a)$ is represented in Fig.~\ref{fig2}. The unit-cell is composed
of four CMONS molecules arranged according to the monoclinic
$C_{\scriptscriptstyle{c}}$ crystalline space group
\cite{CMONScristallo}. The macroscopic polarization responses from
mono-crystalline nanocrystals are dependent on their unit-cell
orientation $\Omega$. The TPF optical response involves the
molecular fourth order $\gamma$ and second order $\alpha$
susceptibility tensors, which participate respectively to the
two-photon absorption and the one-photon emission processes. The
TPF intensity analyzed in the $I$ polarization direction is
proportional to \cite{LeFloch}:
\begin{eqnarray}
\mathcal{I}_I^{\textrm{TPF}}(\Omega)=\!\alpha_{\scriptscriptstyle{II}}(\Omega)\!\!\!\!\sum_{J,K,L,M}\!\!
\gamma_{\scriptscriptstyle{JKLM}}(\Omega)\overline{
E_{\scriptscriptstyle{J}} E_{\scriptscriptstyle{K}}
E_{\scriptscriptstyle{L}} E_{\scriptscriptstyle{M}}}
\label{eq:one}
\end{eqnarray}
where the $I,J,K,L,M$ indexes span the macroscopic frame ($X,Y$),
and $\overline{(...)}$ denotes photodetection temporal averaging.
The $E_{\scriptscriptstyle{I}}$ components are the incident field
polarization projections on each macroscopic axis. The SHG optical
response stems from the molecular third order $\beta$
susceptibility tensor. The resulting intensity analyzed in the $I$
polarization direction for a mono-crystalline structure originates
from the coherent addition of induced nonlinear polarizations from
each molecule in the crystal, and is therefore proportional to:
\begin{eqnarray}
\mathcal{I}_I^{\textrm{SHG}}(\Omega)=\!\!\sum_{J,K,L,M}\!
\beta_{\scriptscriptstyle{IJK}}(\Omega)\beta_{\scriptscriptstyle{ILM}}(\Omega)
\overline{E_{\scriptscriptstyle{J}}E_{\scriptscriptstyle{K}}
E_{\scriptscriptstyle{L}}E_{\scriptscriptstyle{M}}} \label{eq:two}
\end{eqnarray}
Using Eqs.~(\ref{eq:one}),~(\ref{eq:two}) and in-plane rotation of
the incident polarization, each of the macroscopic coefficients
$\alpha_{\scriptscriptstyle{II}}$,
$\beta_{\scriptscriptstyle{JKL}}$ and
$\gamma_{\scriptscriptstyle{IJKL}}$ can be deduced. These
coefficients depend on the orientation angle $\Omega$ and on the
unit-cell susceptibility components
$\alpha_{\scriptscriptstyle{ij}}$,
$\gamma_{\scriptscriptstyle{jklm}}$ and
$\beta_{\scriptscriptstyle{ijk}}$ with $(i,j,k)=$ (a,b,c). For any
$t$ tensor with components $t_{\scriptscriptstyle{I...N}}$
expressed in the macroscopic frame, the $\Omega$ dependence of
these coefficients can be readily calculated from a projection law
$t_{\scriptscriptstyle{I...N}}(\Omega)=\sum_{i,...,n}
t_{\scriptscriptstyle{i...n}}\cos(i,I)(\Omega)...\cos(n,N)(\Omega)
$, where the $\cos(i,I)$ functions of ($\theta,\phi,\psi$) express
the unit-cell frame projection in the macroscopic frame.
Consequently, the exploration of polarization responses finally
relies on the knowledge of the microscopic $\alpha$ and $\gamma$
tensor coefficients for TPF detection, and of the $\beta$ tensor
coefficients for SHG. At the molecular scale, we modelled a single
CMONS molecule by a rod-like system of susceptibility components
denoted $\alpha_{\scriptscriptstyle{uu}}^{\textrm{cmons}}$,
$\beta_{\scriptscriptstyle{uuu}}^{\textrm{cmons}}$ and
$\gamma_{\scriptscriptstyle{uuuu}}^{\textrm{cmons}}$, $u$ defining
the CMONS fundamental dipole direction in the molecular frame
\cite{Wang}. The unit-cell tensor coefficients are constructed
from the addition of each molecular contribution accounting for
their respective orientation in the unit cell, according to the
oriented gas model. This model neglects possible contributions
from intermolecular interactions in the susceptibility
calculations \cite{NLOcrystals}. Numerical values of the resulting
$\alpha$, $\beta$ and $\gamma$ tensor coefficients in the
unit-cell (a,b,c) frame are given in the table of Fig.~\ref{fig2},
the few non-vanishing remaining coefficients being consistent with
the unit-cell symmetry. The off-diagonal tensor coefficients are
the signature of the multipolar symmetry of the CMONS
nanocrystals, which strongly influences the polarization
dependence of the macroscopic optical responses.

\begin{figure}[!h]
\includegraphics[width=7cm]{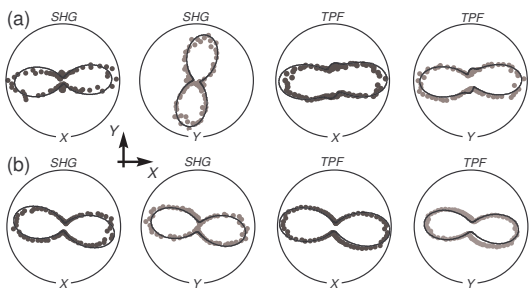}
\caption{\label{fig3} TPF and SHG polarization responses of two
mono-crystalline nanocrystals in the $X$ and $Y$ analyzed
polarization directions (indicated at the bottom of the circles).
The data are displayed with similar sizes, although they are of
different amplitudes. The fits with the mono-crystalline model
give the following orientation parameters within a $\pm 5$\% error
range: (a): ($50^{\circ}$,$60^{\circ}$,$70^{\circ}$), (b):
($80^{\circ}$,$80^{\circ}$,$110^{\circ}$). These angles are
defined within the $[0,\pi]$ angular range. Note that the
combinations
($\theta$,$\phi$,$\psi$),($\theta$,$\phi$+$\pi$,$\psi$) and
($\theta$,$\phi$,$\psi$+$\pi$) give identical polarization
responses.}
\end{figure}

Fig.~\ref{fig3} shows a typical example of signals from two
isolated nanocrystals, exhibiting various TPF and SHG polarization
responses analyzed in the $X$ and $Y$ directions. These signals
can be simultaneously fitted with the previous model, which is a
clear evidence of mono-crystalline structures. Moreover, the fit
of both TPF and SHG features permits the retrieval of the
orientation parameters of such nanocrystals, using
Eqs~(\ref{eq:one}) and~(\ref{eq:two}) \footnote{The fits account
for the large aperture of the microscope objective, as well as the
slight ellipticity and dichroism resulting from reflection of the
incident IR beam on the dichroic mirror, characterized separately
by ellipsometry.}. This example shows a typical situation where
two nanocrystals of different orientations exhibit similar TPF
polarization responses, with on the contrary very different SHG
responses. This originates from the distinct tensorial nature of
these two optical processes: TPF involves even order tensors and
is sensitive to the axial order terms of the molecular
distribution (symmetric contributions), whereas SHG is sensitive
to odd order parameters (non-centrosymmetric contributions). This
complementarity is therefore the key point in the determination of
the 3D orientation of isolated nanocrystals \footnote{The 3D
orientation determination would however not be possible in the
rare case of rod-like symmetry
($C_{\scriptscriptstyle{{\infty}v}}$) crystals, for which the
projection of the unit-cell in the sample plane is of the same
symmetry as the system itself.}. In comparison, traditional
fluorescence microscopy only measures the in-plane orientation of
the emitters, while the access to out-of plane contributions
requires the use of complex optical field geometries or variable
incident angle measurements \cite{SingleMolecule,Wild}.
In the present measurements, the determination of the three
($\theta,\phi,\psi$) angles is furthermore unambiguous. Indeed, a
slight change of one of the three Euler angles (by typically
$\pm\,5-10^{\circ}$) would clearly modify differently the TPF and
SHG polarization responses. Moreover another type of analysis,
which consists of using the experimental TPF and SHG anisotropy
ratios in order to determine the unit-cell orientation, gives
similar results by ruling-out any solution that does not
correspond to polar plots such as those shown in Fig.~\ref{fig3}.

\begin{figure}[!h]
\includegraphics[width=7cm]{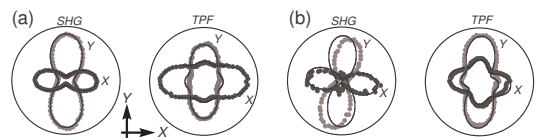}
\caption{\label{fig4} TPF and SHG polarization responses of two
non-mono-crystalline nanocrystals in the $X$ and $Y$ analyzed
polarization directions. (a): a nearly-isotropic nanocrystal. (b):
a poly-crystalline nanocrystal, for which the polar plots fits are
obtained from the incoherent addition of two nanocrystals of
orientations ($70^{\circ}$,$160^{\circ}$,$60^{\circ}$) and
($70^{\circ}$,$55^{\circ}$,$65^{\circ}$). This combination of
angles is not the unique solution for the fit.}
\end{figure}

It is visible from the measurements on mono-crystalline structures
of Fig.~\ref{fig3} that the TPF polarization responses in the $X$
and $Y$ analysis directions have the same shape, although they are
of different amplitudes. This effect is also predicted
mathematically by the model. A departure from this main
characteristic is therefore evidence of a non mono-crystalline
phase which contains at least two or more nanocrystals. Such
behavior, as represented in Fig.~\ref{fig4}, is typically observed
for about 30\% of the measured nanocrystals. The TPF and SHG
features of Fig.~\ref{fig4}(a) are identical and have a ($X,Y$)
axial symmetry for crossed polarization excitations. Such
symmetry, fixed by the detection polarization directions, is a
signature of incoherent emission from an isotropic orientational
distribution of emitters. This effect, which has been observed in
Rayleigh scattering experiments in solutions \cite{HLS}, has been
only moderately investigated so far in the study of nanoparticles
\cite{NLOgold}. In the case of the CMONS nanoparticles, an
incoherent SHG emission is made possible by their sub-wavelength
size, contrary to micrometer-scale particles for which the SHG
efficiency is expected to depend strongly on the particle size
when it approaches the coherence length. Such an effect can occur
from the nano-scale assembly of randomly oriented SHG-active
nanocrystals formed during the crystallization process. An
incoherent summation of the responses from $N\!\!>\!\!1$
nanocrystals of different orientations contained in a
sub-wavelength size aggregate can be modelled by the following SHG
or TPF intensities:
\begin{eqnarray}
\mathcal{I}_I^{\textrm{incoherent}}&=& \sum_{N}
\mathcal{I}_I(\Omega_N) \label{eq:five}
\end{eqnarray}
with $\mathcal{I}(\Omega_N)$ being the SHG or TPF intensities from
a nanocrystal of given orientation $\Omega_N$ in the aggregate.
Computing the SHG and TPF responses with an increasing amount of
randomly generated $\Omega_N$ orientations, the completely
symmetric features of Fig.~\ref{fig4}(a) start appearing from an
incoherent summation over $N\!>\!10$ nanocrystals. Less-symmetric
features are the signature of an intermediate situation between
mono-crystalline and isotropic assemblies of nanocrystals
originating from a sum over 2 to 10 nanocrystals. In
Fig.~\ref{fig4}(b), a reasonable fit is obtained for $N$=$2$
nanocrystals in the aggregate, having different relative
orientations. Our observations therefore allow the distinction
between mono-crystalline and more complex poly-crystalline
structures formed by either a nearly-isotropic aggregate of
nanocrystals, or by a small number ($N\!\!<\!\!10$ for the current
analysis) of nanocrystals with sub-wavelength sizes.

In conclusion, the combination of TPF and SHG in two-photon
nonlinear polarized microscopy has allowed the examination of
organic nanoparticles, with the possibility to clearly distinguish
mono-crystalline structures from poly-crystalline arrangements. We
have also shown that a quantitative model of the  TPF and SHG
polarization responses enables the 3D orientation of
mono-crystalline nanocrystals imbedded in an amorphous host
matrix. This method can be extended to a broad variety of complex
structures. The study of both polarized SHG and TPF at the
nanoscale level can be applied to investigate local field effects
and local orientation of molecules under electrical or optical
fields perturbation. It can furthermore be extended to
orientational tracking of optical markers in complex environments.

The authors thank Pr. J.-F. Nicoud (IPCMS Strasbourg) for his
assistance in the selection and preparation of the chromophore, V.
De Beaucoudrey for experimental contributions, and R. Pansu for
fruitful discussions. This work is partially supported by an ACI
"Jeune Chercheur" grant from Minist\`{e}re de la Recherche,
France.

%\bibliographystyle{apsrev}
%\bibliography{bibcmons}
%\end{document}

\end{document}